%% file: B2G-12-010_temp.tex
\pdfoutput=1

\documentclass[11pt,twoside,a4paper,cmspaper,final,collab]{cms-tdr}
\def\svnVersion{250002}\def\cmsCernNo{2014-011}\def\cmsCernDate{\today}\def\cmsMessage{Published in the Journal of High Energy Physics as \href{http://dx.doi.org/10.1007/JHEP05(2014)108}{\doi{10.1007/JHEP05(2014)108.}}}

\begin{document}\cmsNoteHeader{B2G-12-010}

\hyphenation{had-ron-i-za-tion}
\hyphenation{cal-or-i-me-ter}
\hyphenation{de-vices}

\RCS$Revision: 232179 $
\RCS$HeadURL: svn+ssh://svn.cern.ch/reps/tdr2/papers/B2G-12-010/trunk/B2G-12-010.tex $
\RCS$Id: B2G-12-010.tex 232179 2014-03-17 17:12:42Z narain $
\cmsNoteHeader{B2G-12-010} 

\newcommand{\PWprL}{\ensuremath{\cmsSymbolFace{W}_\cmsSymbolFace{L}^{\prime}}\xspace}
\newcommand{\PWprR}{\ensuremath{\cmsSymbolFace{W}_\cmsSymbolFace{R}^{\prime}}\xspace}
\newcommand{\PWprMix}{\ensuremath{\cmsSymbolFace{W}_\cmsSymbolFace{LR}^{\prime}}\xspace}
\renewcommand{\PWpr}{\ensuremath{\cmsSymbolFace{W}^{\prime}}}
\newcommand{\PWprPlus}{\ensuremath{\cmsSymbolFace{W}^{\prime+}}\xspace}
\newcommand{\aL}{a^{\mathrm{L}}\xspace}
\newcommand{\aR}{a^{\mathrm{R}}\xspace}

\title{Search for $\PWpr\to\cPqt\cPqb$ decays in the lepton + jets final state in pp collisions at $\sqrt{s}=8\TeV$}

\date{\today}

\abstract{
Results are presented from a search for the production of a heavy gauge boson \PWpr\ decaying into a top and a bottom quark, using a data set collected by the CMS experiment at $\sqrt{s}=8\TeV$ and corresponding to an integrated luminosity of 19.5\fbinv.  Various models of \PWpr-boson production are studied by allowing for an arbitrary combination of left- and right-handed couplings. The analysis is based on the detection of events with a lepton ($\Pe, \Pgm$), jets, and missing transverse energy in the final state. No evidence for \PWpr-boson production is found and 95\% confidence level upper limits on the production cross section times branching fraction are obtained. For \PWpr\ bosons with purely right-handed couplings, and for those with left-handed couplings assuming no interference effects, the observed 95\% confidence level limit is $M(\PWpr)>2.05\TeV$.  For \PWpr\ bosons with purely left-handed couplings, including interference effects, the observed 95\% confidence level limit is $M(\PWpr)>1.84\TeV$. The results presented in this paper are the most stringent limits published to date.
}

\hypersetup{%
pdfauthor={CMS Collaboration},%
pdftitle={Search for W' to tb decays in the lepton + jets final state in pp collisions at sqrt(s) = 8 TeV},%
pdfsubject={CMS},%
pdfkeywords={CMS, physics, B2G, exotica, Wprime}}

\maketitle 

\section{Introduction}

Massive charged gauge bosons, generically referred to as {\PWpr}, are predicted by various extensions of the standard model (SM)~\cite{doi:10.1146/annurev.nucl.55.090704.151502,PhysRevD.64.035002,PhysRevD.64.065007,PhysRevD.53.5258,PhysRevD.11.566}.  Searches for {\PWpr} bosons at the Large Hadron Collider (LHC) have been conducted in the lepton-neutrino, diboson, and light-quark final states~\cite{CMSWprimeLNu8TeV,CMSWprimeLNu7TeV,ATLASWprimeLNu7TeV,CMSWZtag7TeV,CMSVZ7TeV,CMSWprimeWZ7TeV,ATLASDiboson7TeV,ATLASWprimeWZ7TeV,CMSDijet8TeV,ATLASDijet7TeV}. While the most stringent limits come from the searches in the leptonic final states ($\PWpr \to \ell\nu$ where $\ell$ is a charged lepton), these constraints do not apply to {\PWpr} bosons with purely right-handed couplings if the mass of the hypothetical right-handed neutrino is larger than a few \GeVns~\cite{leftrightatlhc}. Dedicated searches for {\PWpr} bosons with purely right-handed couplings have been performed by the CMS and ATLAS Collaborations assuming the mass of the right-handed neutrino is less than the mass of the {\PWpr} boson~\cite{cmswprimer,atlaswprimer}. Searches for right-handed {\PWpr} bosons that decay to a quark final state such as ${\PWprPlus} \to \cPqt\cPaqb$ (or charge conjugate) make no assumptions regarding the mass of the right-handed neutrino and are thus complementary to searches in the leptonic channels. Furthermore, the decay chain ${\PWpr} \to \cPqt\cPqb$, $\cPqt \to \cPqb{\PW} \to \cPqb\ell\nu$  is in principle fully reconstructable,  thereby leading to observable resonant mass peaks even in the case of broad {\PWpr} resonances. In addition, because of the presence of leptons in the final state, it is easier to suppress the continuum multijet background for this decay chain than for a generic $\PWpr\to \Pq\Pq^\prime$ decay.  Finally, in some models the {\PWpr} boson may couple more strongly to fermions of the third generation than to fermions of the first and second generations~\cite{Muller1996345,Malkawi1996304}.  Thus the $\PWpr \to \cPqt\cPqb$ decay is an important channel in the search for {\PWpr} bosons.

Experimental searches for $\PWpr \to \cPqt\cPqb$ decays have been performed at the Tevatron~\cite{CDF:2009,D0:2010,D0Wprime} and at the LHC~\cite{Wprimetb2011, Aad:2012ej}. The CMS search at $\sqrt{s}=7$\TeV~\cite{Wprimetb2011} set the best present mass limit in this channel of 1.85\TeV for {\PWpr} bosons with purely right-handed couplings. If the {\PWpr} boson has left-handed couplings, interference between $\PWpr \to \cPqt\cPqb$ and SM single-top-quark production via $\PW \to \cPqt\cPqb$ can contribute as much as 5--20\%  of the total {\PWpr} rate, depending on the \PWpr\ mass and couplings~\cite{Boos:2006xe}. This interference effect was taken into account in the CMS search. The CMS analysis also set constraints on an arbitrary set of left- and right-handed couplings of the {\PWpr} boson.

This Letter describes the first $\PWpr \to \cPqt\cPqb$ search in pp collisions at $\sqrt{s}=8$\TeV and uses data collected by the CMS experiment corresponding to an integrated luminosity of 19.5\fbinv. For a {\PWpr} boson with a mass of 2\TeV, the production cross section at $\sqrt{s}=8$\TeV is larger by approximately a factor of two compared to $\sqrt{s}=7$\TeV~\cite{Boos:2006af}. The data set used in this analysis corresponds to an integrated luminosity that is approximately a factor of four larger than that in the $\sqrt{s}=7$\TeV analysis. Following the approach of the earlier publication~\cite{Wprimetb2011}, we analyse events with an electron (\Pe) or muon ($\mu$), jets, and missing transverse energy (\ETmiss) for an arbitrary combination of left- and right-handed couplings.

\section{CMS detector}
\label{sec:CMS}
The central feature of the CMS detector is a superconducting solenoid of 6\unit{m} internal diameter, providing a magnetic field of 3.8\unit{T}. Located within the superconducting solenoid volume are a silicon pixel and strip tracker, a lead tungstate crystal electromagnetic calorimeter (ECAL), and a brass and scintillator hadron calorimeter (HCAL). Muons are identified and measured in gas-ionisation detectors embedded in the outer steel
magnetic flux-return yoke of the solenoid. The detector is subdivided into a cylindrical barrel and endcap disks on each side of the interaction point. Forward calorimeters complement the
coverage provided by the barrel and endcap detectors. A more detailed description of the CMS detector can be found elsewhere~\cite{:2008zzk}.

The CMS experiment uses a right-handed coordinate system, with the origin at the nominal interaction point, the $x$ axis pointing to the centre of the LHC ring, the $y$ axis pointing up (perpendicular to the plane of the LHC ring), and the $z$ axis along the anticlockwise-beam direction. The polar angle $\theta$ is measured from the positive $z$ axis and the azimuthal angle $\phi$ is measured in radians in the $x$-$y$ plane. The pseudorapidity $\eta$ is defined as $\eta = -\ln[\tan(\theta/2)]$.

The ECAL energy resolution for electrons with transverse energy $\ET\approx45$\GeV from $\cPZ \to \Pe \Pe$ decays is better than 2\% in the central region of the ECAL barrel $(\abs{\eta} < 0.8)$, and is between 2\% and 5\% elsewhere.
The inner tracker measures charged particles within the pseudorapidity range $\abs{\eta}< 2.5$. It provides an impact parameter resolution of ${\sim}15\mum$ and a transverse momentum (\pt) resolution of about 1.5\% for 100\GeV particles. Matching muons to tracks measured in the silicon tracker results in a relative transverse momentum resolution for muons with $20 <\pt < 100\GeV$ of 1.3--2.0\% in the barrel and better than 6\% in the endcaps. The \pt resolution in the barrel is better than 10\% for muons with \pt up to 1\TeV~\cite{Chatrchyan:2012xi}.

A particle-flow (PF) algorithm~\cite{pf, pf2} combines the information from all CMS subdetectors to
identify and reconstruct the individual particles emerging from all vertices: charged hadrons,
neutral hadrons, photons, muons, and electrons. These particles are then used to reconstruct
the \ETmiss (defined as the modulus of the negative transverse momentum vector sum of all measured particles), jets, and to quantify lepton isolation.
The PF jet energy resolution is typically 15\% at 10\GeV, 8\% at 100\GeV, and 4\% at 1\TeV, to be compared to about 40\%, 12\%, and 5\% obtained when the calorimeters alone are used for jet clustering.

\section{\label{sec:MCmodel}Signal and background modelling}
The $\PWpr \to \cPqt\cPqb \to \ell\nu\mathrm{bb}$ decay is characterized by the presence of a high-\pt isolated lepton, significant \ETmiss associated with the neutrino, and at least two high-\pt b-jets (jets resulting from the fragmentation and hadronization of b quarks). Monte Carlo (MC) techniques are used to model the {\PWpr} signal and SM backgrounds capable of producing this final state.

\subsection{Signal modelling}
The signal modelling is identical to that in Ref.~\cite{Wprimetb2011} and uses the following lowest order effective Lagrangian to describe the interaction of the $\PWpr$ boson with SM fermions:
\begin{equation}
\mathcal{L} = \frac{V_{f_if_j}}{2\sqrt{2}} g_w \overline{f}_i\gamma_\mu
\bigl( \aR_{f_if_j} (1+{\gamma}^5) + \aL_{f_if_j}
(1-{\gamma}^5)\bigr) {\PWpr}^{\mu} f_j + \text{h.c.},
\end{equation}
where $\aR_{f_if_j}, \aL_{f_if_j}$ are the right- and left-handed couplings of the {\PWpr} boson to fermions $f_i$ and $f_j$, $g_w = e/(\sin{\theta_W})$ is the SM weak coupling constant and $\theta_\PW$ is the weak mixing angle; $V_{f_if_j}$ is the  Cabibbo--Kobayashi--Maskawa matrix element if the fermion $f$ is a quark, and $V_{f_if_j}=\delta_{ij}$ if it is a lepton, where $\delta_{ij}$ is the
Kronecker delta and $i,\ j$ are the generation numbers. For our search we consider models where $0 \leq a^{\mathrm{L,R}}_{f_if_j} \leq 1$.
For a SM-like \PWpr\ boson, $\aL_{f_if_j}=1$ and $\aR_{f_if_j}=0$.

We simulate ${\PWpr}$ bosons with mass values ranging from 0.8 to 3.0\TeV. The \textsc{singletop} MC generator~\cite{Boos:2006af} is used, which simulates electroweak top-quark production processes based on the complete set of tree-level Feynman diagrams calculated by the {\COMPHEP} package~\cite{Boos:2004kh}. Finite decay widths and spin correlations between resonance state production and subsequent decay are taken into account.  The factorisation scale is set to the ${\PWpr}$-boson mass for the generation of
the samples and the computation of the leading-order (LO) cross section. The LO cross section is scaled to next-to-leading order (NLO) using a $K$ factor of 1.2 based on Refs.~\cite{Sullivan:2002jt,Duffty:2012rf}.
In order to ensure that the NLO rates and shapes of relevant distributions are reproduced, the \textsc{ singletop}  generator includes NLO corrections, and normalisation and matching between various partonic subprocesses are performed. The top-quark mass is chosen to be 172.5\GeV and the CTEQ6M~\cite{Pumplin:2002vw} parton distribution functions (PDF) are used. The uncertainty in the cross section is
about 8.5\% and includes contributions from the uncertainties in the renormalisation and factorisation scales (3.3\%), PDFs (7.6\%),
$\alpha_s$ (1.3\%), and the top-quark mass (${<}1$\%).

We produce the following sets of signal samples:
\begin{itemize}
\item{${\PWprL}$ with $\aL_{\cPqu\cPqd}=\aL_{\cPqc\cPqs}=\aL_{\cPqt\cPqb}=1~\text{and}~\aR_{\cPqu\cPqd}=\aR_{\cPqc\cPqs}=\aR_{\cPqt\cPqb}=0$ }
\item{${\PWprR}$ with $\aL_{\cPqu\cPqd}=\aL_{\cPqc\cPqs}=\aL_{\cPqt\cPqb}=0~\text{and}~\aR_{\cPqu\cPqd}=\aR_{\cPqc\cPqs}=\aR_{\cPqt\cPqb}=1$ }
\item{${\PWprMix}$ with $\aL_{\cPqu\cPqd}=\aL_{\cPqc\cPqs}=\aL_{\cPqt\cPqb}=1~\text{and}~\aR_{\cPqu\cPqd}=\aR_{\cPqc\cPqs}=\aR_{\cPqt\cPqb}=1$ }
\end{itemize}

The ${\PWprL}$ bosons couple to the same fermion multiplets as the SM W boson.
As a consequence, there will be interference between $s$-channel
tb production via a W boson and via a ${\PWprL}$ boson.
These two processes therefore cannot be generated
separately. Thus the ${\PWprL}$ and ${\PWprMix}$ samples include
SM $s$-channel tb production including its interference  with the ${\PWprL}$ signal.
Production of a tb final state via a ${\PWprR}$ boson does not interfere with
tb production via a W boson and therefore the ${\PWprR}$
sample only includes $\PWpr$ production.

The ${\PWprR}$ boson can only decay leptonically if there is a right-handed neutrino
$\nu_{\mathrm{R}}$  of sufficiently small mass, $M(\nu_{\mathrm{R}})$, so
that $M(\nu_{\mathrm{R}})+M(\ell)<M({\PWpr})$.
If the mass of the right-handed neutrino is too large, ${\PWprR}$
bosons can only decay to $\mathrm{q}\overline{\mathrm{q}}^{\prime}$
final states, leading to different branching fractions for
the ${\PWprR}\to \cPqt\cPqb$ decay than for the ${\PWprL}\to \cPqt\cPqb$
decay. In the absence of interference between the SM W boson and the ${\PWpr}$
 boson, and if there is a light right-handed neutrino, there is no practical difference
for our search between  ${\PWprL}$ and ${\PWprR}$ bosons.

\subsection{Background modelling}
The $\ttbar$, W+jets, single-top-quark ($s$-channel, $t$-channel, and tW associated production), $\cPZ/\gamma^*$+jets, and diboson (WW) background contributions are estimated from simulation, with corrections to the shape and normalisation derived from data.

The $\ttbar$,  $\PW$+jets, and $\cPZ/\gamma^*$+jets background processes are generated with {\MADGRAPH} 5.1~\cite{madgraph5}.
The {\ttbar} background is normalized to the next-to-NLO (NNLO) cross section~\cite{Czakon:2013goa}. The SM single-top-quark backgrounds are estimated using samples generated
with {\POWHEG}~\cite{powheg}, normalized to an approximate NNLO cross section~\cite{Kidonakis:2012rm}.
For the ${\PWprR}$ search, $s$-channel,  $t$-channel, and tW single-top-quark events
are considered as backgrounds. Because of interference between $\PWpr$ and $s$-channel single-top-quark production, in the analysis
for  ${\PWprL}$ and ${\PWprMix}$ bosons only the $t$-channel and the tW processes contribute to the background.
The diboson (WW) background is generated with \PYTHIA 6.424~\cite{Sjostrand:2006za}. Instrumental background due to a jet misidentified as an isolated lepton was studied using a sample of QCD multijet events simulated with {\PYTHIA} and was found to be negligible after the final selection.

\subsection{Simulation}
For all simulated samples, \PYTHIA tune Z2*~\cite{UEPAS} is used for parton showering, hadronisation, and simulation of the underlying event. The \PYTHIA and {\MADGRAPH} backgrounds use the CTEQ6L1 PDFs, and the \POWHEG backgrounds use the CTEQ6M PDFs~\cite{Pumplin:2002vw}. The resulting events are processed with the full \GEANTfour~\cite{GEANT} simulation of the CMS detector. The additional proton-proton interactions in each beam crossing (pileup) are modelled by superimposing extra minimum-bias interactions onto simulated events, with the distribution of the number of pileup interactions matching that in data.

\section{Object and event preselection}
\label{sec:selection}

The analysis relies on the reconstruction of electrons, muons, jets, and \MET. Candidate events are required to pass an isolated electron (muon) trigger with a {\pt} threshold of 27 (24)\GeV and to have at least one reconstructed pp interaction vertex. In the offline selection, exactly one electron (muon) is required to be
within the region of $\abs{\eta}<$ 2.5 (2.1).  Additionally, the barrel/endcap transition region,  $1.44 < \abs{\eta}< 1.56$, is excluded for electrons. Electrons and muons are required to satisfy $\pt > 50$\GeV and a series of identification and isolation criteria. Electron candidates are selected using shower shape information, the quality of the track, the matching between the track and the electromagnetic cluster, the fraction of total cluster energy in the HCAL, and the amount of activity in the surrounding regions of the tracker and calorimeters. Events are removed whenever the electron is found to
originate from a converted photon. The track associated with a muon candidate is required to have at least one pixel hit, hits in at least six layers of the inner tracker, at least one hit in the muon detector, and a good quality fit with $\chi^2/\mathrm{d.o.f.} < 10$. Both electrons and muons are separated from jets by requiring $\Delta R(\text{jet},\ell)=\sqrt{\smash[b]{(\Delta\eta)^2+(\Delta\phi)^2}}>0.3$. Additionally, the cosmic ray background is effectively eliminated by requiring
the transverse impact parameter of the muon with respect to the beam spot to
be less than 2\unit{mm}. Electrons (muons) are required to have PF based relative isolation, $I_\text{ rel}$, less than 0.10 (0.12). The quantity $I_\text{rel}$ is defined as the sum of the transverse momenta of all additional reconstructed particle candidates inside a cone
 around the electron (muon) in $(\eta,\phi)$ of $\Delta R  < 0.3$ (0.4), divided by the
$\pt$ of the electron (muon). An event-by-event correction is applied to the computation of the lepton isolation in order to account for the effect of pileup. Events containing a second lepton with looser identification and isolation requirements are also rejected. Scale factors, derived from comparing the efficiencies measured in data and simulation using $\cPZ\to \ell\ell$ events, are obtained for lepton identification and isolation as a function of lepton \pt and $\eta$. These are applied as corrections to the simulated events.

Jets are clustered using the anti-\kt algorithm~\cite{antikt} with a distance parameter of $R=0.5$ and are required to
satisfy $\pt > 30\GeV$ and $\abs{\eta}< 2.4$. At least two jets are required in the event with the
highest-$\pt$ (leading) jet $\pt > 120\GeV$ and the second leading jet $\pt > 40\GeV$. The jet $\pt$ in the simulated samples is smeared to account for the better jet energy resolution observed in the simulation compared to data~\cite{CMSJESJER}. Jet energy corrections are applied to correct for residual non-uniformity and non-linearity of the detector response. Jet energies are also corrected by subtracting the average contribution from pileup interactions~\cite{Fastjet1,Fastjet2}.

The final state of the $\PWpr \to \cPqt\cPqb$ decay includes two b quarks; therefore at least one of the two leading jets is required to be tagged as a b-jet.  We use the combined secondary vertex tagger with the medium operating point~\cite{CMS:Btagging}. Data-to-simulation scale factors for the b-tagging efficiency and the light-quark or gluon (udsg) jet mistag rate are applied on a jet-by-jet basis to all b-jets,
c-jets, and udsg jets in the simulated events. Scale factors are also applied to \PW+jets events in which a b, c, or udsg jet is produced in association with the \PW~boson, in order to bring the data and simulation yields into agreement. The procedure used is identical to the one described in Ref.~\cite{Wprimetb2011}.  Based on lepton + jets samples with various jet multiplicities, W+b and W+c corrections are derived~\cite{PhysRevD.84.092004}. To account for differences between the lepton + jets topology and the topology considered here, additional W+udsg and W+b/c corrections are derived from two background-dominated event samples, one without any b-tagged jets and one without any b-tagging requirement. These corrections are then applied to the simulated \PW+jets events.   We find that the W+b, W+c, and W+udsg contributions need to be corrected by an overall factor of 1.21, 1.66, and 0.83, respectively.  These corrections agree within their uncertainties with the corresponding corrections derived in Ref.~\cite{Wprimetb2011}.

Finally, the \MET is required to exceed 20\GeV in both the electron and muon samples in order to reduce the QCD multijet background.

\section{Data analysis}
\label{sec:invmass}

The distinguishing feature of a \PWpr\ signal is a narrow
resonance structure in the tb invariant-mass spectrum.  The tb invariant mass is reconstructed from the combination of the charged lepton, the neutrino, the jet which gives the
best top-quark mass reconstruction, and the highest-$\pt$ jet in the event that is not associated with the top quark.
The $x$ and $y$ components of the neutrino momentum are obtained from the missing transverse energy. The
$z$ component is calculated by constraining the  invariant mass of the lepton-neutrino pair to the $\PW$-boson mass (80.4\GeV).
This constraint leads to a quadratic equation in  $p_z^{\nu}$. In the case
of two real solutions, both of the solutions are used
to reconstruct the $\PW$-boson candidates. In the case of complex solutions, the
real part is assigned to $p_z^{\nu}$ and the imaginary part is forced to zero by relaxing the $\PW$-boson mass constraint and
recomputing $\pt^{\nu}$.  The $\pt^{\nu}$ solution that gives the invariant mass of the lepton-neutrino pair closest to
80.4\GeV is chosen, resulting in a single $\PW$-boson candidate. Top-quark candidates are then reconstructed using the
$\PW$-boson candidate(s) and all of the selected jets in the event, and the
top-quark candidate with mass closest to 172.5\GeV is chosen. The {\PWpr}-boson candidate
is obtained by combining the best top-quark candidate
with the highest-$\pt$ jet, excluding the one used for the best top-quark candidate. For a 2.0\TeV $\PWprR$ boson, this procedure assigns the correct jets from the $\PWpr$ decay 83\% of the time.

Since the $\PW$+jets process is one of the major backgrounds for the {\PWpr} signal process
(see Table~\ref{yields}), a study is performed to check that the shape of the $\PW$+jets mass distribution is well-modelled by the simulation. This cross-check utilizes the fact that events
that have no b-tagged jets, but satisfy all other selection criteria, are
expected to originate predominantly from $\PW$+jets events. The purity of $\PW$+jets events for this
control sample is greater than 85\%.
The shape of the $\PW$+jets background is obtained by subtracting the backgrounds from sources other than $\PW$+jets from the distributions in data.
The resulting invariant-mass distribution is compared to the distribution from the $\PW$+jets MC sample with zero b-tagged jets.
The difference between
the distributions is included as a systematic uncertainty in the shape of the
$\PW$+jets background. Using simulated events, the $\PW$+jets background was verified to be independent
of the number of b-tagged jets by comparing the mass distribution with zero b-tagged jets with that obtained
by requiring one or more b-tagged jets.

Measurements of the top-quark differential cross sections have shown that the top-quark $\pt$ distribution is not properly modelled in simulated events~\cite{CMSTopDiff}.  We therefore reweight the $\ttbar$ sample using an empirical function of the
generated top quark and anti-quark $\pt$
determined from studies of the $\ttbar$ differential cross section. Residual differences with respect to the unweighted
distribution are  taken into account as a systematic uncertainty in the $\ttbar$ background prediction.  We check the applicability
of these weights to our kinematic region by defining a control region in data that is dominated by $\ttbar$ events.  The control region
is defined by the following requirements, which are designed to ensure small ($\lesssim$2\%) potential signal contamination:
$N_{\text{jets}} \geq 4$, the total number of b-tagged jets (including jets with $\pt$ values less than those of the two leading jets) $N_{\text{b-tags}} \geq 2$, and $ 400 < M(\cPqt\cPqb) < 750\GeV$.
We perform a fit to the ratio of data to expected background events for the top-quark $\pt$ distribution using a Landau function and
reweight the events in the simulated $\ttbar$ sample using the result of
the fit.  This method gives results that are consistent with the generator-level reweighting procedure.

\begin{figure}[!h!tbp]

\centering
\includegraphics[width=0.75\textwidth]{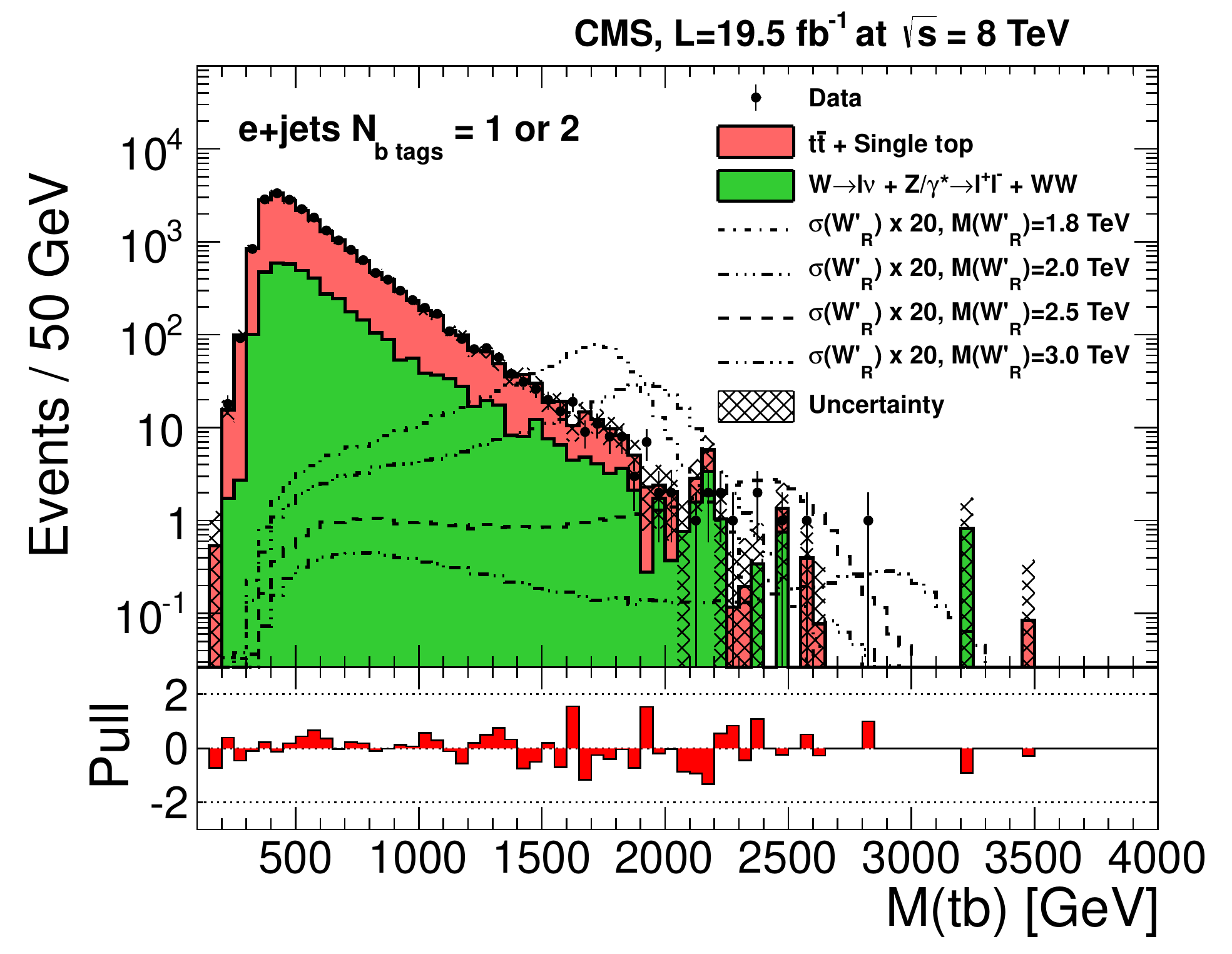}
\includegraphics[width=0.75\textwidth]{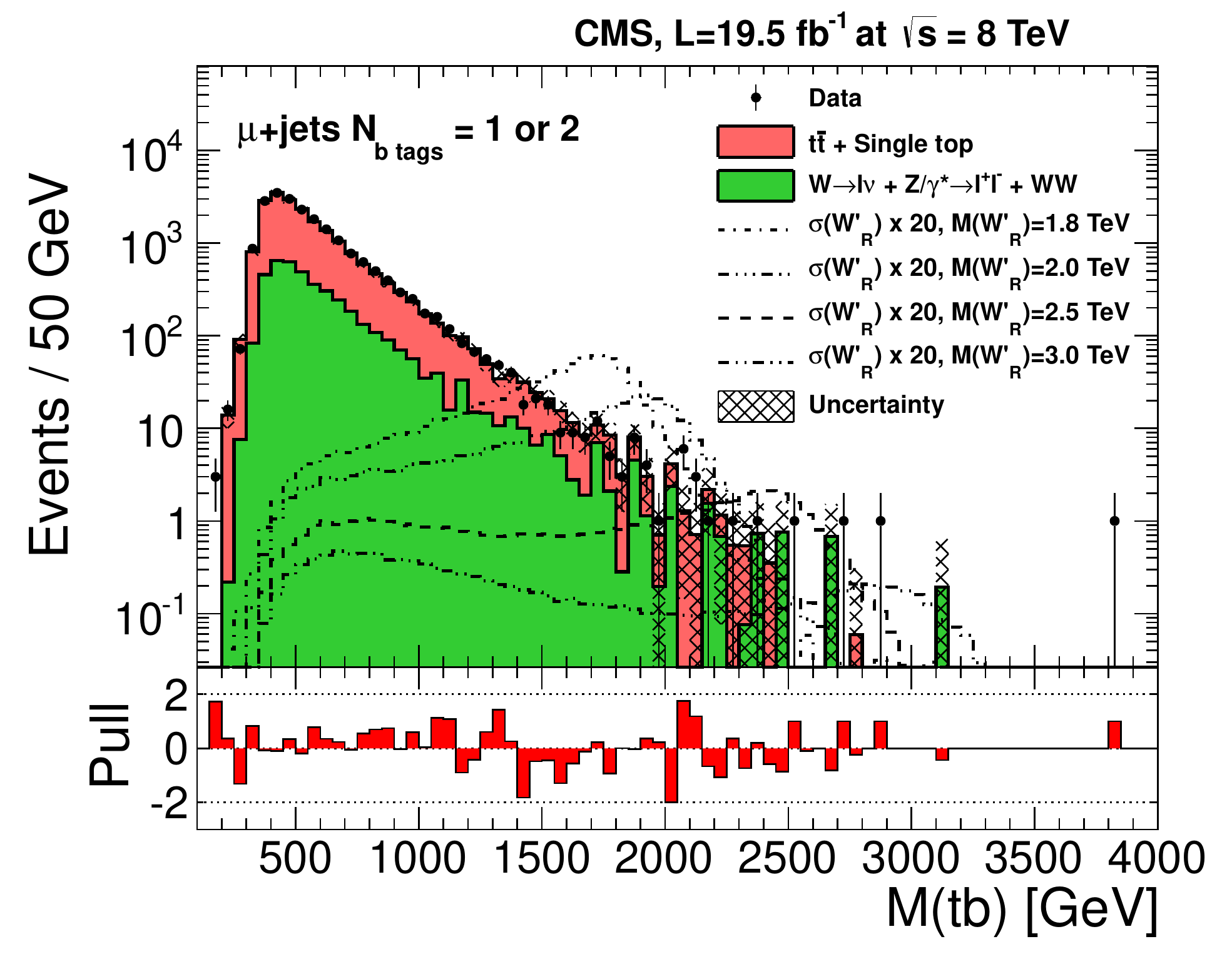}

\caption{
The reconstructed invariant-mass distribution of the {\PWpr}-boson candidates after the final selection.
Events with electrons (muons) are shown on the left (right) panel
for data, background and four different ${\PWprR}$ signal mass hypotheses (1.8, 2.0, 2.5, and 3.0\TeV).
All events are required to have one or both of the two leading jets tagged as b-jets.
The hatched bands represent the total normalisation uncertainty in the predicted backgrounds.
The pull is defined as the difference between the observed data yield and the predicted background, divided by the uncertainty.
For these plots it is assumed that $M(\nu_{\mathrm{R}}) \ll M(\PWprR)$ and for the purpose of illustration the expected yields for the $\PWprR$ signal samples are scaled by a factor of 20.
\label{Fig:data-bkg}}
\end{figure}

Figure~\ref{Fig:data-bkg} shows the reconstructed tb invariant-mass distribution
obtained from data and from simulated {\PWpr} signal samples with four different mass
values ($M(\PWpr)= 1.8$, 2.0, 2.5, and 3.0\TeV). Also shown are the dominant background
contributions. The distributions are shown after the preselection described in Section~\ref{sec:selection}, as well
as three final selection criteria which are imposed to improve
the signal-to-background discrimination: the $\pt$ of the selected top-quark candidate $\pt^{\cPqt}>$ 85\GeV, the $\pt$ of the vector sum of the two leading
jets $\pt^{\text{jet1,jet2}}>140\GeV$, and the mass of the selected top-quark candidate
with $130\GeV < M(\cPqt) < 210\GeV$. The distributions are shown separately for the electron and muon samples, for events which have one or both of the two leading jets tagged as b-jets.
The number of events remaining with one and two b-tagged jets after the preselection and final selection are listed in Table~\ref{yields}.  The yields measured in data and those predicted from simulation agree within the statistical and systematic uncertainties, which are described in the following section.

\begin{table*}[!h!tb]
\centering
\topcaption{
     Number of selected data, signal, and background  events.
    For the background samples, the number of expected events is computed
    corresponding to an integrated luminosity of 19.5\fbinv.
    The final two columns for each sample include the following selections: $\pt^{\cPqt}>85\GeV$, $\pt^{\text{jet1,jet2}}>140\GeV$, $130 < M(\cPqt) <210\GeV$. The combined statistical and systematic uncertainty on the total background prediction is also shown. The standard model $s$-channel tb process contributes to the background only in the search for $\PWprR$ bosons owing to its interference with the $\PWprL \to$ tb process.  The number of events for the $\PWprL$ signal takes into account the interference with the SM $s$-channel tb process.
\label{yields}
        }
\small
\footnotesize
\begin{tabular}{l|c|c|c|c|c|c|c|c}
\hline
    & \multicolumn{8}{c}{Number of selected events} \\ \hline
  &  \multicolumn{4}{c|}{ Electron sample }  & \multicolumn{4}{c}{ Muon sample } \\
  &  \multicolumn{2}{c|}{Preselection} & \multicolumn{2}{c|}{Final selection} & \multicolumn{2}{c|}{Preselection}  & \multicolumn{2}{c}{Final selection}  \\
 Process &  1 b-tag   & 2 b-tags  & 1 b-tag  & 2 b-tags & 1 b-tag  & 2 b-tags  & 1 b-tag & 2 b-tags \\ \hline
 \multicolumn{9}{l}{\bf{Signal:} } \\ \hline
$M(\PWprR)$ = 1.8 TeV & 45.2 & 12.7 & 32.2 & 9.3 & 38.0 & 10.8 & 26.3 & 7.7  \\
$M(\PWprR)$ = 2.0 TeV & 20.9 & 5.6 & 14.6 & 4.0 & 17.5 & 4.7 & 11.8 & 3.2  \\
$M(\PWprR)$ = 2.5 TeV & 3.5 & 0.9 & 2.3 & 0.6 & 3.0 & 0.8 & 1.8 & 0.5  \\
$M(\PWprR)$ = 3.0 TeV & 0.8 & 0.3 & 0.5 & 0.2 & 0.7 & 0.2 & 0.4 & 0.2  \\
$M(\PWprL)$ = 1.8 TeV & 143.0 & 60.9 & 57.1 & 19.7 & 148.8 & 63.7 & 58.1 & 19.5  \\
$M(\PWprL)$ = 2.0 TeV & 125.2 & 57.9 & 44.7 & 17.8 & 128.3 & 61.0 & 45.7 & 18.1 \\
$M(\PWprL)$ = 2.5 TeV & 115.8 & 58.6 & 38.4 & 17.2 & 122.3 & 62.6 & 41.6 & 17.7 \\
$M(\PWprL)$ = 3.0 TeV & 121.3 & 58.1 & 41.0 & 16.7 & 126.6 & 64.4 & 42.2 & 17.9  \\ \hline \hline
\multicolumn{9}{l}{\textbf{Background:}}     \\ \hline
$\ttbar$  & 34561 & 7888 & 12383 & 1639 & 35349 & 8191 & 12610 & 1650  \\
$s$-channel (tb) & 175 & 93 & 58 & 28 & 196 & 102 & 63 & 32  \\
$t$-channel (tqb) & 2113 & 357 & 710 & 108 & 2275 & 373 & 747 & 114  \\
tW-channel & 2557 & 362 & 847 & 107 & 2645 & 372 & 861 & 113  \\
$\PW(\to\ell\nu)$+jets & 19970 & 563 & 3636 & 99& 19697 & 679 & 3704 & 62  \\
$\cPZ/\gamma^*(\to\ell\ell$)+jets  & 1484 & 83 & 260 & 10 & 1497 & 73 & 275 & 17  \\
WW & 205 & 9 & 47 & 3 & 219 & 7 & 47 & 2  \\ \hline
\textbf{Total bkg.} & 61065 & 9357 & 17942 & 1993 & 61877 & 9797 & 18307 & 1991 \\
 & $\pm$6188 & $\pm$1504 & $\pm$2514 & $\pm$399 & $\pm$6098 & $\pm$1524 & $\pm$2488 & $\pm$400  \\ \hline\hline
\textbf{Data}  & 63050 & 9646 & 18175 & 2063 & 62955 & 9865 & 18558 & 2081  \\ \hline
\textbf{Total bkg. / Data}  & 0.969 & 0.970 & 0.987 & 0.966 & 0.983 & 0.993 & 0.986 & 0.957 \\
 & $\pm$0.10 & $\pm$0.16 & $\pm$0.14 & $\pm$0.19 & $\pm$0.10 & $\pm$0.15 & $\pm$0.13 & $\pm$0.19 \\ \hline
\end{tabular}
\end{table*}

\section{\label{sec:sys}Systematic uncertainties}
The systematic uncertainties that are relevant for this analysis fall into two categories: (i)~uncertainties in the total event yield and (ii)~uncertainties that impact both the shape and the total event yield of the distributions. The first category includes uncertainties in the total integrated luminosity of the data sample (2.6\%)~\cite{LumiUncertainty}, lepton reconstruction and identification efficiencies (1\%), trigger modelling (1--2\%), and the theoretical {\ttbar} cross section (8\%).

The second category includes the uncertainty from the jet energy scale and resolution,
and from the b-tagging and the mis-tagging efficiency scale factors.
For the \PW+jets samples, uncertainties relating to the extraction of the light- (13\%) and heavy-flavour (15\%) scale factors from data are also included~\cite{CMS:Btagging}.
As discussed in the previous section, additional uncertainties are assigned relating to the \PW+jets background shape and to
the top quark $\pt$ spectrum.
The variation of the renormalisation and factorisation scale $Q^2$ used in the
strong coupling constant $\alpha_s(Q^2)$, and the jet-parton matching scale
uncertainties in the MLM scheme~\cite{MLM} are evaluated for the {\ttbar} background sample.
These uncertainties are evaluated by raising and lowering the corresponding parameters by one standard deviation (or in the case of the renormalisation and factorisation scale $Q$ and the jet parton matching scale by a factor 2 and 0.5), and repeating the analysis.

\section{Results}
The $\PWpr$-boson mass distribution
observed in the data and the prediction
for the total expected background agree within statistical and systematic uncertainties (see Table~\ref{yields} and Fig.~\ref{Fig:data-bkg}).
We set upper limits on the {\PWpr}-boson production cross section for
different {\PWpr}-boson masses.

\subsection{Cross section limits}

The limits are computed using a Bayesian approach with a
flat prior on the signal cross section with the \textsc{theta} package~\cite{theta-stat}.
In order to reduce the bin-by-bin statistical uncertainty in the predicted event yields obtained from the simulated samples, we bin the invariant-mass distribution using one bin from 100 to 300\GeV, 17 bins of 100\GeV width from 300 to 2000\GeV, and two
additional bins from 2000 to 2200\GeV and from 2200 to 4000\GeV. Four categories are defined according to the lepton flavor
(electron or muon) and b-tag
multiplicity (one or two b-tagged jets) to improve the sensitivity of the analysis. The resulting distributions serve as the
inputs to the limit setting procedure, and the limit is based on the
posterior probability defined by using all categories simultaneously.
A binned likelihood is used to calculate upper limits on the signal
production cross section times total leptonic branching fraction:
$\sigma (\Pp\Pp\to \PWpr ) \times {\mathcal{B}}(\PWpr \to \cPqt\cPqb \to \ell\nu\cPqb\cPqb$),
where $\ell = \Pe/\mu/\tau$.  The search is sensitive to the $\PWpr \to \cPqt\cPqb \to \tau\nu\cPqb\cPqb$ decay mode
if the tau subsequently decays to an electron or muon. Therefore $\tau \to \Pe/\mu$ events are included in the signal and background
estimations of the electron and muon samples, respectively.
The limit computation accounts for the effects of systematic uncertainties (discussed in Section \ref{sec:sys}) in the normalisation and shape of the invariant-mass distributions, as well as for
statistical fluctuations in the background templates.
Expected limits on the production cross section for each ${\PWprR}$-boson mass are also computed
as a measure of the sensitivity of the analysis.

\begin{figure}[!htbp]
  \centering     \includegraphics[width=0.49\textwidth]{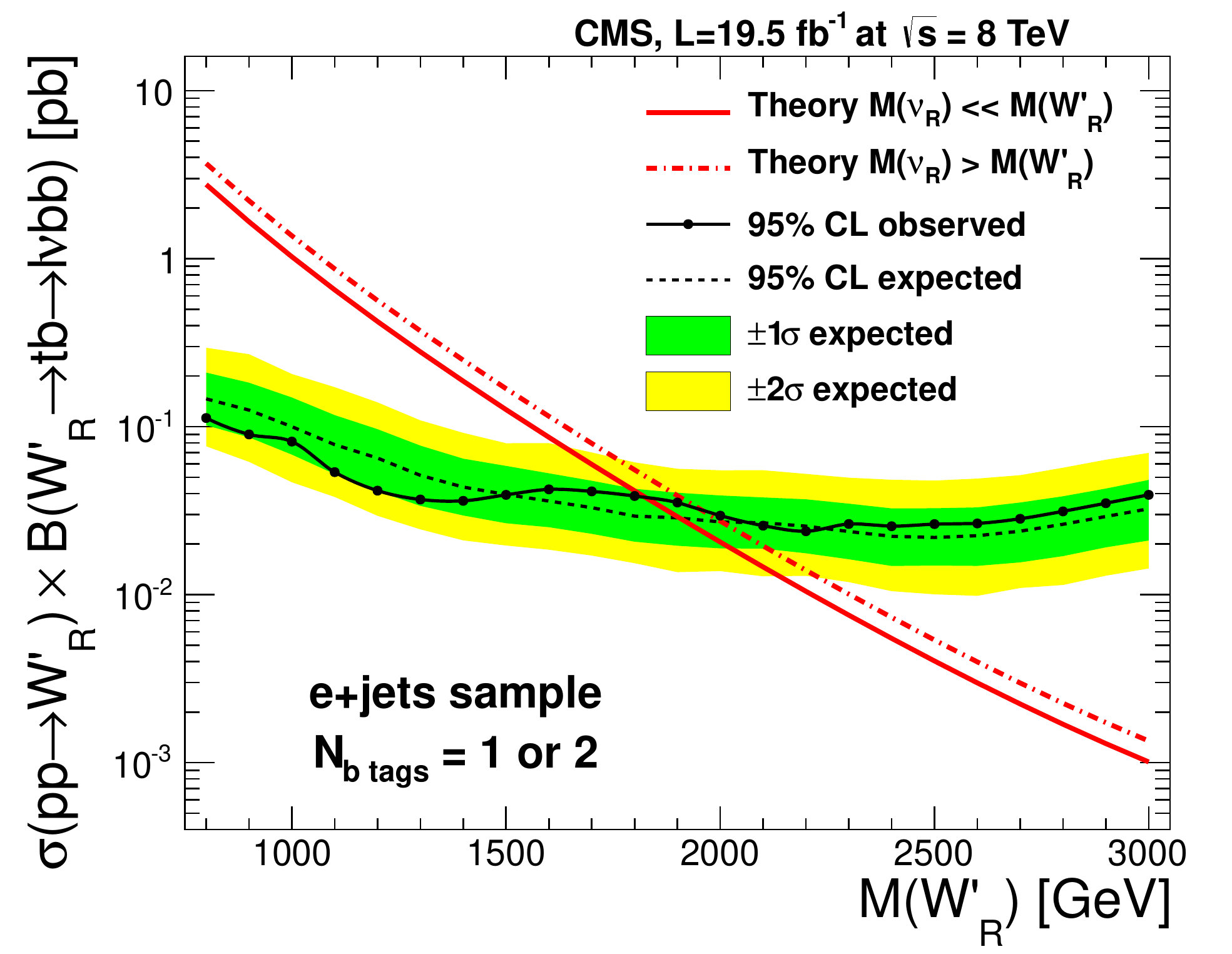}
  \includegraphics[width=0.49\textwidth]{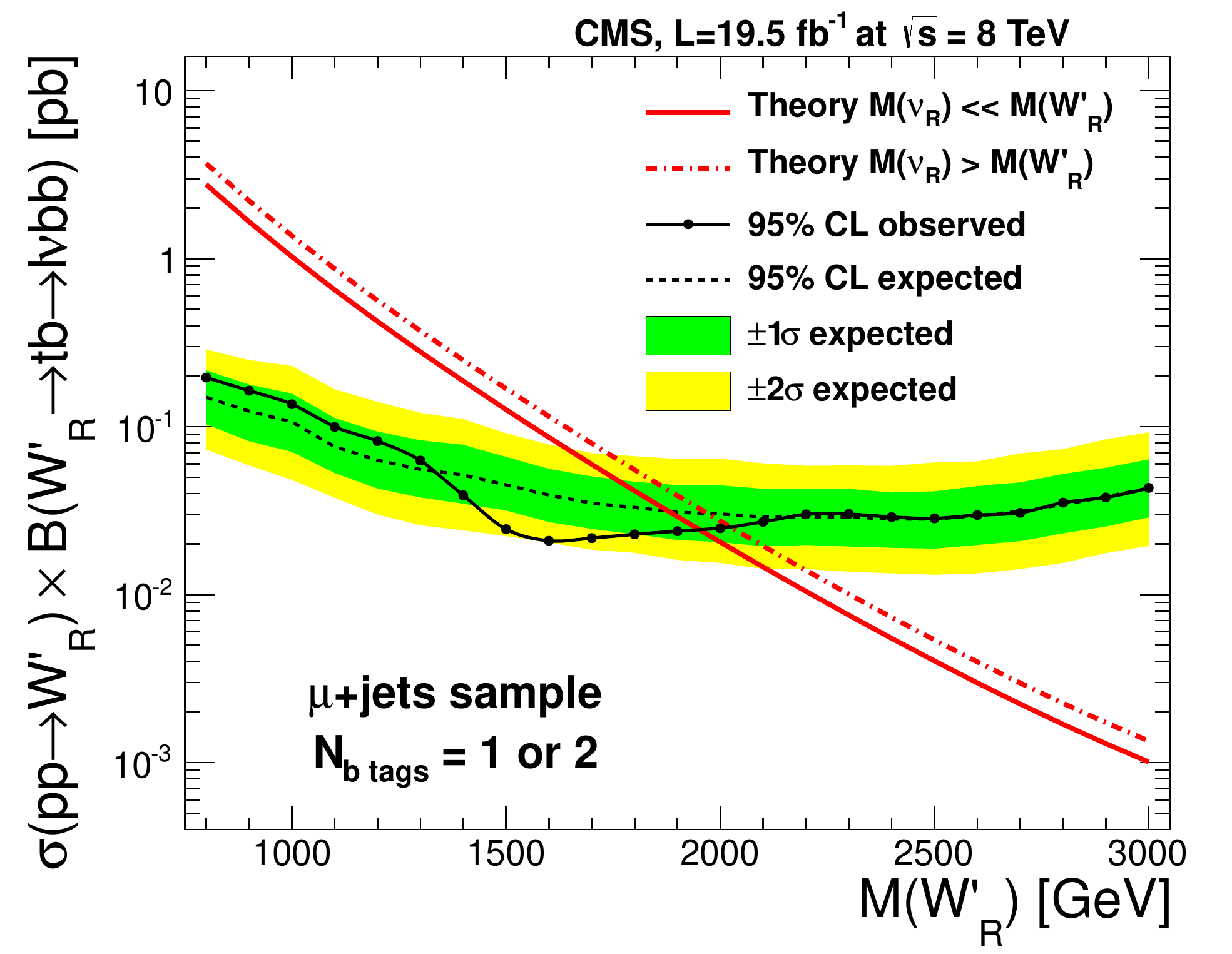}
  \includegraphics[width=0.65\textwidth]{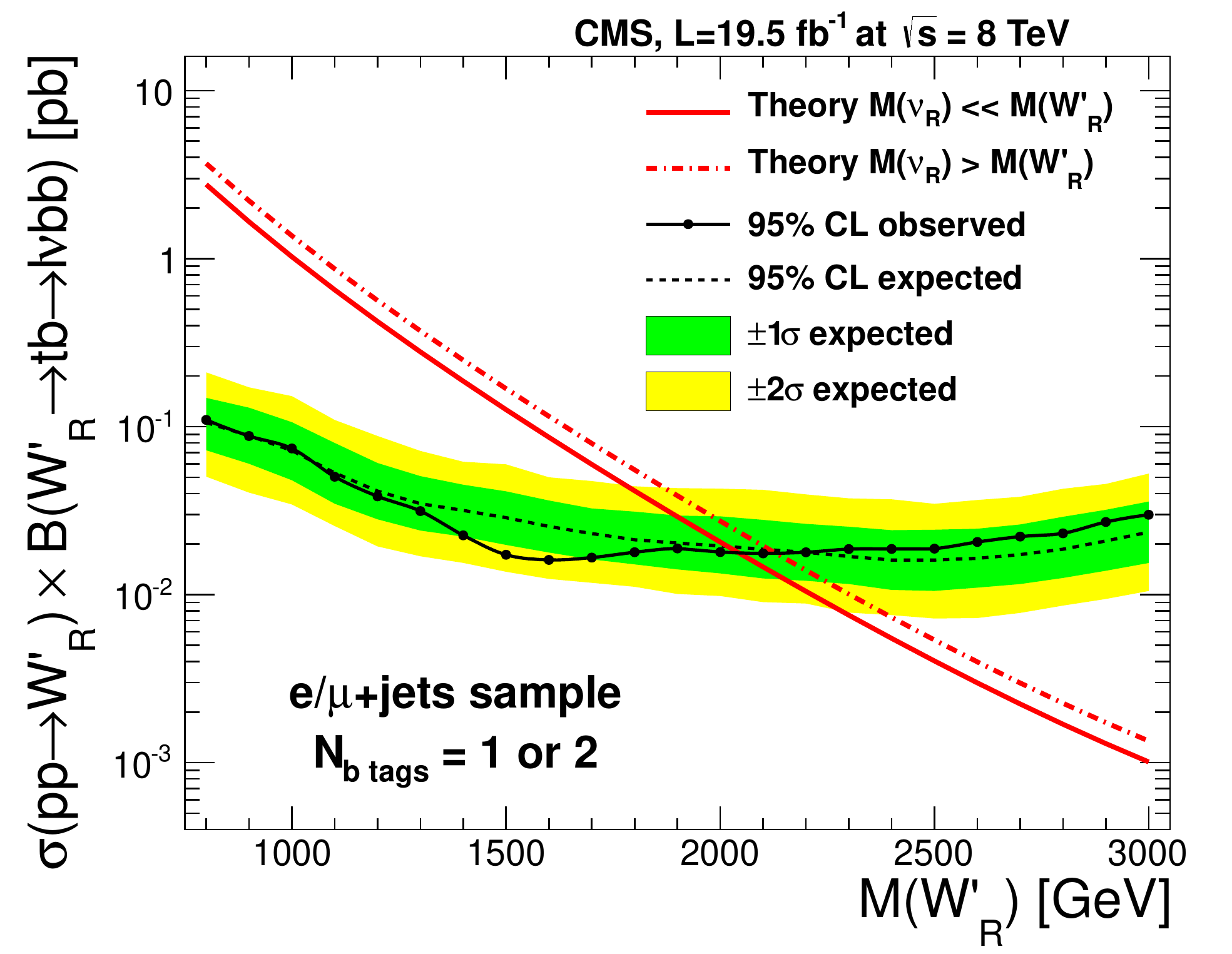}
  \caption{The expected (dashed black line) and observed (solid black line) 95\% CL upper limits on the production cross section of right-handed {\PWpr} bosons obtained for the electron sample (top left), muon sample (top right), and their combination (bottom) along with the ${\pm}1\sigma$ and ${\pm} 2\sigma$ uncertainty in the expected exclusion limit.  The theoretical cross section times branching fraction for right-handed \PWpr-boson production $\sigma(\Pp\Pp \to \PWprR) \times \cPqb(\PWprR \to \cPqt\cPqb \to \ell\nu\cPqb\cPqb)$, where $\ell = \Pe/\mu/\tau$, is shown as a solid (dot-dashed) red line, when assuming light (heavy) right-handed neutrinos.
  \label{fig:limit-right}}
\end{figure}

In Fig.~\ref{fig:limit-right}, 
the solid black line denotes the observed limit and the red lines
represent the predicted theoretical cross section times leptonic branching fractions.
The lower mass limit is defined by the mass value corresponding to the intersection of the observed upper limit on the production cross section times leptonic branching fraction with the
theoretical prediction. For \PWpr\ bosons with right-handed couplings to fermions the observed (expected) limit is 2.05 (2.02)\TeV at 95\% confidence level (CL). These limits also apply to a left-handed $\PWpr$ boson when no interference with the SM is taken into account. Assuming heavy right-handed neutrinos ($M(\nu_{\mathrm{R}}) > M(\PWpr)$),  the observed (expected) limit is 2.13 (2.12)\TeV at 95\% CL.

\subsection{Limits on coupling strengths\label{Shat-generalCoupling}}

The effective Lagrangian given by Eq.~(1) can be analysed for arbitrary combinations of
left-handed or right-handed coupling strengths~\cite{Wprimetb2011}.
The cross section for single-top-quark production in the presence of a {\PWpr}
boson for any set of coupling
values can be written in terms of the cross sections of our signal MC samples,
${\sigma}_{\mathrm{L}}$ for purely  left-handed  couplings $(\aL,\aR)=(1,0)$,
${\sigma}_{\mathrm{R}}$ for purely  right-handed couplings $(\aL,\aR)=(0,1)$,
${\sigma}_{\mathrm{LR}}$ for mixed  couplings $(\aL,\aR)=(1,1)$,
and  ${\sigma}_{\mathrm{SM}}$ for SM couplings $(\aL,\aR)=(0,0)$. It is given by:
\begin{equation}\begin{aligned}\label{eq:xsec}
{\sigma} &= {\sigma}_{\mathrm{SM}} + \aL_{\cPqu\cPqd}\aL_{\cPqt\cPqb}
\left({\sigma}_{\mathrm{L}} - {\sigma}_{\mathrm{R}} - {\sigma}_{\mathrm{SM}} \right)  \\
         &+ \left(\left(\aL_{\cPqu\cPqd} \aL_{\cPqt\cPqb}\right)^2
          +        \left(\aR_{\cPqu\cPqd} \aR_{\cPqt\cPqb}\right)^2\right) {\sigma}_{\mathrm{R}} \\
         &+ \frac{1}{2}\left(\left(\aL_{\cPqu\cPqd} \aR_{\cPqt\cPqb}\right)^2
          +                   \left(\aR_{\cPqu\cPqd} \aL_{\cPqt\cPqb}\right)^2\right)
\left( {\sigma}_{\mathrm{LR}} - {\sigma}_{\mathrm{L}} - {\sigma}_{\mathrm{R}}  \right).
\end{aligned}\end{equation}
Note that for pure ${\PWprR}$ production this reduces to the sum of SM $s$-channel tb and ${\PWprR}$ production.
For pure ${\PWprL}$ or ${\PWprMix}$ production this reduces to the cross section of the ${\PWprL}$ or the ${\PWprMix}$
sample which already includes SM $s$-channel tb production and its interference with $\PWpr$ production.

We assume that the couplings to first-generation quarks, $a_{\cPqu\cPqd}$, that are important for the production of the
{\PWpr} boson, and the couplings to third-generation quarks, $a_{\cPqt\cPqb}$,
that are important for the decay of the {\PWpr} boson, are equal.
The event samples are combined according to Eq.~(\ref{eq:xsec}) to give the predicted invariant-mass distributions for each value of $\aL$ and $\aR$.

\begin{figure*}[!h!tb]
\centering
 \includegraphics[width=0.48\textwidth]{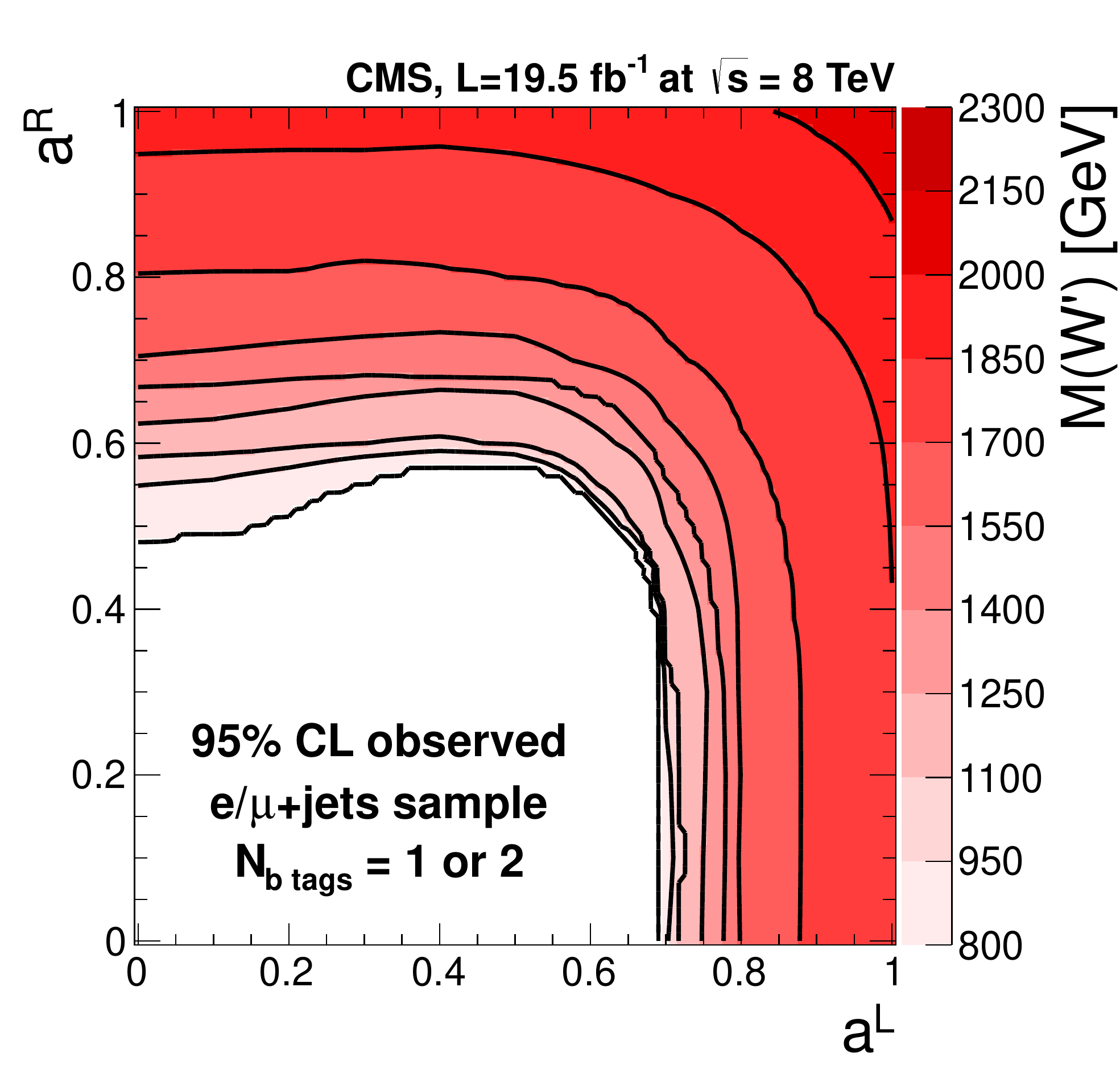}
 \includegraphics[width=0.48\textwidth]{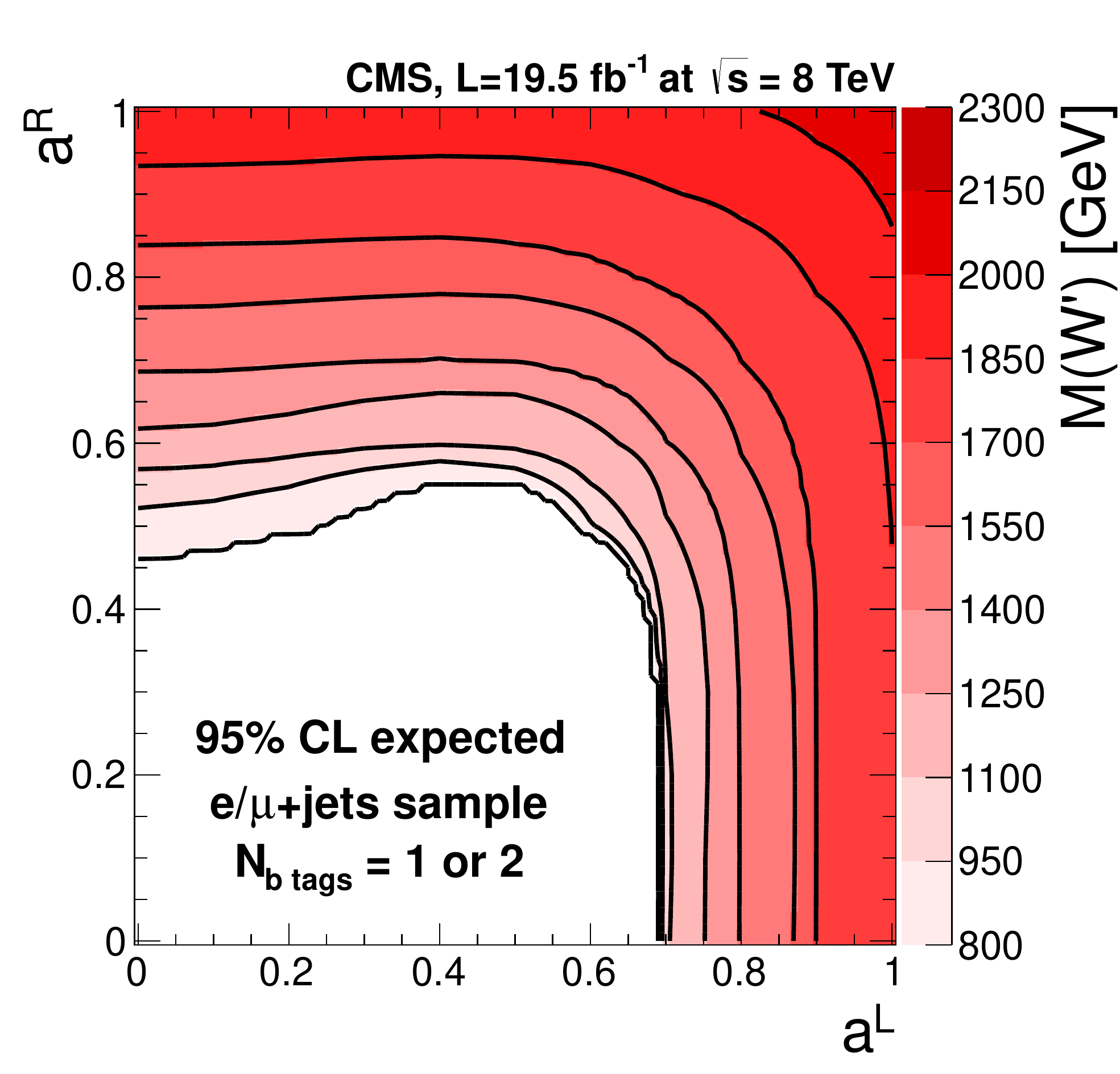}
 \caption{
  Contour plots of  $M(\PWpr)$ in the ($\aL,\aR$) plane
  for which the 95\% CL cross section limit equals the predicted
  cross section for the combined $\Pe,\mu$+jets sample.
 The left (right) panel represents the observed (expected) limits.
 The colour axis represents the value of $M(\PWpr)$ in \GeVns. The solid black lines are isocontours of {\PWpr}-boson mass, plotted in 150\GeV intervals and starting from 800\GeV.
 \label{fig:disc_limit}}
 \end{figure*}

We vary both $\aL$ and $\aR$ in the range (0,1) with a step size of 0.1, for each $M({\PWpr})$. For each of these
combinations of $\aL$, $\aR$,  and $M({\PWpr})$, we determine the
expected and observed 95\% CL upper limits on the cross section and compare them to the
corresponding theoretical prediction. If the limit is below the theoretical
prediction, this point in ($\aL,\aR,M({\PWpr})$) space is excluded. Figure~\ref{fig:disc_limit} shows
the excluded {\PWpr}-boson mass for each point in the ($\aL,\aR$) plane.  The observed (expected) mass limit
for a {\PWpr} boson with only left-handed couplings, including interference with the SM, is 1.84 (1.84)\TeV.

\section{Summary}

We have performed a search for a $\PWpr$ boson in the tb decay
channel using a data set corresponding to an integrated luminosity of 19.5\fbinv of pp collisions collected by the CMS detector at $\sqrt{s}=8\TeV$. No evidence for the presence of a $\PWpr$ boson is found, and 95\% confidence level upper limits on $\sigma (\Pp\Pp \to \PWpr) \times \mathcal{B}(\PWpr \to \cPqt\cPqb \to \ell \nu \cPqb\cPqb)$ are set. We compare
our measurement to the theoretical prediction for the
cross section to determine the lower limit on the mass of the $\PWpr$ boson.
For $\PWpr$ bosons with right-handed couplings to fermions
(and for left-handed couplings to fermions, when assuming no interference effects)
 the observed (expected) limit is 2.05 (2.02)\TeV at 95\% confidence level.  In the case with heavy right-handed neutrinos ($M(\nu_{\mathrm{R}}) > M(\PWprR)$),  the observed (expected) limit is 2.13 (2.12)\TeV at 95\% confidence level.  For a $\PWpr$ boson with only left-handed couplings, including interference effects, the observed (expected) limit is 1.84 (1.84)\TeV at 95\% confidence level.
We also set constraints on the $\PWpr$ gauge coupling independent of their chiral structure. The results presented in this paper are the most stringent limits obtained to date.

\section*{Acknowledgements}

We congratulate our colleagues in the CERN accelerator departments for the excellent performance of the LHC and thank the technical and administrative staffs at CERN and at other CMS institutes for their contributions to the success of the CMS effort. In addition, we gratefully acknowledge the computing centres and personnel of the Worldwide LHC Computing Grid for delivering so effectively the computing infrastructure essential to our analyses. Finally, we acknowledge the enduring support for the construction and operation of the LHC and the CMS detector provided by the following funding agencies: BMWF and FWF (Austria); FNRS and FWO (Belgium); CNPq, CAPES, FAPERJ, and FAPESP (Brazil); MES (Bulgaria); CERN; CAS, MoST, and NSFC (China); COLCIENCIAS (Colombia); MSES and CSF (Croatia); RPF (Cyprus); MoER, SF0690030s09 and ERDF (Estonia); Academy of Finland, MEC, and HIP (Finland); CEA and CNRS/IN2P3 (France); BMBF, DFG, and HGF (Germany); GSRT (Greece); OTKA and NIH (Hungary); DAE and DST (India); IPM (Iran); SFI (Ireland); INFN (Italy); NRF and WCU (Republic of Korea); LAS (Lithuania); MOE and UM (Malaysia); CINVESTAV, CONACYT, SEP, and UASLP-FAI (Mexico); MBIE (New Zealand); PAEC (Pakistan); MSHE and NSC (Poland); FCT (Portugal); JINR (Dubna); MON, RosAtom, RAS and RFBR (Russia); MESTD (Serbia); SEIDI and CPAN (Spain); Swiss Funding Agencies (Switzerland); NSC (Taipei); ThEPCenter, IPST, STAR and NSTDA (Thailand); TUBITAK and TAEK (Turkey); NASU (Ukraine); STFC (United Kingdom); DOE and NSF (USA).

Individuals have received support from the Marie-Curie programme and the European Research Council and EPLANET (European Union); the Leventis Foundation; the A. P. Sloan Foundation; the Alexander von Humboldt Foundation; the Belgian Federal Science Policy Office; the Fonds pour la Formation \`a la Recherche dans l'Industrie et dans l'Agriculture (FRIA-Belgium); the Agentschap voor Innovatie door Wetenschap en Technologie (IWT-Belgium); the Ministry of Education, Youth and Sports (MEYS) of Czech Republic; the Council of Science and Industrial Research, India; the Compagnia di San Paolo (Torino); the HOMING PLUS programme of Foundation for Polish Science, cofinanced by EU, Regional Development Fund; and the Thalis and Aristeia programmes cofinanced by EU-ESF and the Greek NSRF.

\bibliography{auto_generated}   

\cleardoublepage \appendix\section{The CMS Collaboration \label{app:collab}}\begin{sloppypar}\hyphenpenalty=5000\widowpenalty=500\clubpenalty=5000\input{B2G-12-010-authorlist.tex}\end{sloppypar}
\end{document}

%% file: B2G-12-010-authorlist.tex
\textbf{Yerevan Physics Institute,  Yerevan,  Armenia}\\*[0pt]
S.~Chatrchyan, V.~Khachatryan, A.M.~Sirunyan, A.~Tumasyan
\vskip\cmsinstskip
\textbf{Institut f\"{u}r Hochenergiephysik der OeAW,  Wien,  Austria}\\*[0pt]
W.~Adam, T.~Bergauer, M.~Dragicevic, J.~Er\"{o}, C.~Fabjan\cmsAuthorMark{1}, M.~Friedl, R.~Fr\"{u}hwirth\cmsAuthorMark{1}, V.M.~Ghete, C.~Hartl, N.~H\"{o}rmann, J.~Hrubec, M.~Jeitler\cmsAuthorMark{1}, W.~Kiesenhofer, V.~Kn\"{u}nz, M.~Krammer\cmsAuthorMark{1}, I.~Kr\"{a}tschmer, D.~Liko, I.~Mikulec, D.~Rabady\cmsAuthorMark{2}, B.~Rahbaran, H.~Rohringer, R.~Sch\"{o}fbeck, J.~Strauss, A.~Taurok, W.~Treberer-Treberspurg, W.~Waltenberger, C.-E.~Wulz\cmsAuthorMark{1}
% ...
\vskip\cmsinstskip
% ...
1:~~Also at Vienna University of Technology, Vienna, Austria\\
2:~~Also at CERN, European Organization for Nuclear Research, Geneva, Switzerland\\
% ...
